# Hardware-irrelevant parallel processing system

Xiuting Zou[‡], Shaofu Xu[‡], Anyi Deng, Rui Wang, Weiwen Zou[*]

State Key Laboratory of Advanced Optical Communication Systems and Networks, Intelligent Microwave Lightwave Integration Innovation Center (iMLic), Department of Electronic Engineering, Shanghai Jiao Tong University, Shanghai 200240, China. [‡]These authors contributed equally to this work. [*]Corresponding author: wzou@sjtu.edu.cn

**Abstract**:  Parallel processing technology has been a primary tool for achieving high-speed, high-accuracy, and broadband processing for many years across modern information systems and data processing such as optical and radar, synthetic aperture radar imaging, digital beam forming, and digital filtering systems. However, hardware deviations in a parallel processing system (PPS) severely degrade system performance and pose an urgent challenge. We propose a hardware-irrelevant PPS of which the performance is unaffected by hardware deviations. In this system, an embedded convolutional recurrent autoencoder (CRAE), which learns inherent system patterns as well as acquires and removes adverse effects brought by hardware deviations, is adopted. We implement a hardware-irrelevant PPS into a parallel photonic sampling system to accomplish a high-performance analog-to-digital conversion for microwave signals with high frequency and broad bandwidth. Under one system state, a category of signals with two different mismatch degrees is utilized to train the CRAE, which can then compensate for mismatches in various categories of signals with multiple mismatch degrees under random system states. Our approach is extensively applicable to achieving hardware-irrelevant PPSs which are either discrete or integrated in photonic, electric, and other fields.

In areas of modern information system and data processing, systems such as ultra-wideband radars[1], high-resolution synthetic aperture radar (SAR) imaging[2], phased array radar[3], and digital filtering[4] usually apply parallel processing technology to achieve high-rate sampling, ultra-broadband reception, high-resolution imaging, and quick processing, as well as to reduce the burden of hardware or system cost. However, there are two major hindrances[5-10] for the performance of parallel processing systems (PPSs). First, fixed hardware deviations, such as distinctive attenuations and physical lengths, induce gain and time mismatches. Second, random hardware deviations, such as the variant operation patterns of devices under different environmental conditions, alter channel characteristics and system states. For example, in a fully photonics-based coherent radar system, the different attenuations and delays among the four parallel branches of a photonics-based receiver limit its spurious-free dynamic range[1]; in a multi-channel SAR imaging system, the unavoidable fixed and random hardware deviations among channels severely affect the performance of the signal reconstruction[2].

Traditional approaches[11-16] dealing with time and/or gain mismatches in PPSs are to estimate the volume of mismatches first and then calibrate it. Besides the requirement for knowledge on the input signal, conventional methods are often characterized with complex calculations, noise sensitivity, and additional constraints. Advanced methods such as deep learning[17, 18] are also adopted to calibrate time mismatches. For example, a residual neural network has been trained with a single category of signals collected when the parallel system in one mismatch degree to achieve mismatch compensation. This data-driven approach, however, is only effective on sinusoidal signals with trained mismatch degrees[19]. Thus, it remains challenging to achieve a high-performance PPS through a succinct, generalized, and robust approach which can overcome negative influence resulted from hardware deviations, such as time mismatches, with a low cost.

By means of integrating a PPS and a novel convolutional recurrent autoencoder (CRAE), we propose a hardware-irrelevant PPS of which the performance is unaffected by hardware deviations.  In contrast, the traditional PPS without the CRAE is nominated hardware-relevant PPS. As illustrated in Fig. 1a, the outputs of different PPSs, such as one-dimension waveforms and two-dimension images, are of lower quality than nominal ones due to the bad influence from hardware deviations. The nominal ones are the outputs of the PPSs when there is not any hardware deviation in the systems. We adopt a CRAE (see Fig. 1b), constructed by convolutional neural networks (CNN), recurrent neural networks (RNN), and an autoencoder (AE) structure, to remove adverse effects brought by hardware deviations and to recover the quality of data. The CNN layers strongly extract high-level features from data. The RNN layer is very adaptive to deal with sequence-like objects in that the information of previous neurons affects the output of the latter neurons. The encoder-decoder structure excels in extracting critical information from data and reconstructing the refined version. Convolution and deconvolution are performed to encode and decode data, respectively. Through training CRAE with only one category of signals collected when the PPS is in two fixed hardware deviations (see the green and red curves in Fig. 1b), the inherent patterns of PPSs are implicitly learned; the end-to-end intricate relationship between the low-quality data and the nominal ones is exploited; and thus, the bad influence from hardware deviations is overcome. The trained CRAE is not only effective on untrained categories of signals and untrained fixed hardware deviations (see the pulse and the blue curves in Testing dataset of Fig. 1b) but also robust to random hardware deviations.

## Results

**Implementation of a hardware-irrelevant PPS.** We implement a hardware-irrelevant PPS with an example, namely a hardware-irrelevant parallel photonic sampling system (Supplementary Fig. 1) which consists of a two-channel 20-Gsample/s photonic sampling system and a concrete CRAE (detailed in Methods). It is worth noting that our approach is applicable to various PPSs by virtue of the similarities of system defects brought by hardware deviations. In this parallel photonic sampling system, fixed hardware deviations, such as different physical lengths of the two channels, bring about time mismatches. Random hardware deviations, such as the variant operation patterns of electronic-optical devices on different days, cause variant system states. First, through the parallel photonic sampling system, we obtain mismatched linear frequency modulated (LFM) signals of 1 V with mismatches of 35 ps, 57 ps, and 0 ps (corresponding nominal ones, i.e. reference data). The current random system state is named as State 0. The experimental implementation, data generation, and data acquisition are detailed in Methods. Second, we adopt a minimization algorithm to train the CRAE using LFM signals with mismatches of 35 ps and 57 ps. After training, both the training loss and the testing loss finally converge to a small value about 0.0030 without overfitting (Supplementary Fig. 2). Obtained by calculating the absolute error between the network output and reference data, either the training loss or the testing loss is calculated as an average over the data in the training dataset or the testing dataset, respectively. Thus, we can believe that the end-to-end intricate map between the mismatched signals and the nominal ones is exploited. The trained CRAE is feasible and applicable in mismatch compensation. The details of data processing, the establishment of training and testing datasets, and the training of CRAE are described in Methods. Third, the parallel photonic sampling system is adjusted to be in various mismatch degrees on another day and the random system state is regarded as State 1. Under State 1, the power of the output in the parallel photonic sampling system is almost the same with that under State 0. We obtain several mismatched LFM signals of various amplitudes and Costas frequency modulated signals of 1 V in various mismatch degrees through the parallel photonic sampling system under State 1. Fourth, we verify the capability of the trained CRAE on mismatch compensation and test its generalizability for untrained amplitudes, mismatch degrees, and categories of signals as well as its robustness to random system states. Fifth, we visualize and analyze the trained CRAE for better understanding its working principle. Finally, we adjust the parallel photonic sampling system on another two days and the random system states are regarded as States 2 and 3, respectively. Under State 2, the power of the output in the parallel photonic sampling system is higher than that under State 0. While under State 3, the power of the output in the parallel photonic sampling system is lower. Under States 1-3, we demonstrate the hardware-irrelevant parallel photonic sampling system online with the assistance of LabVIEW.

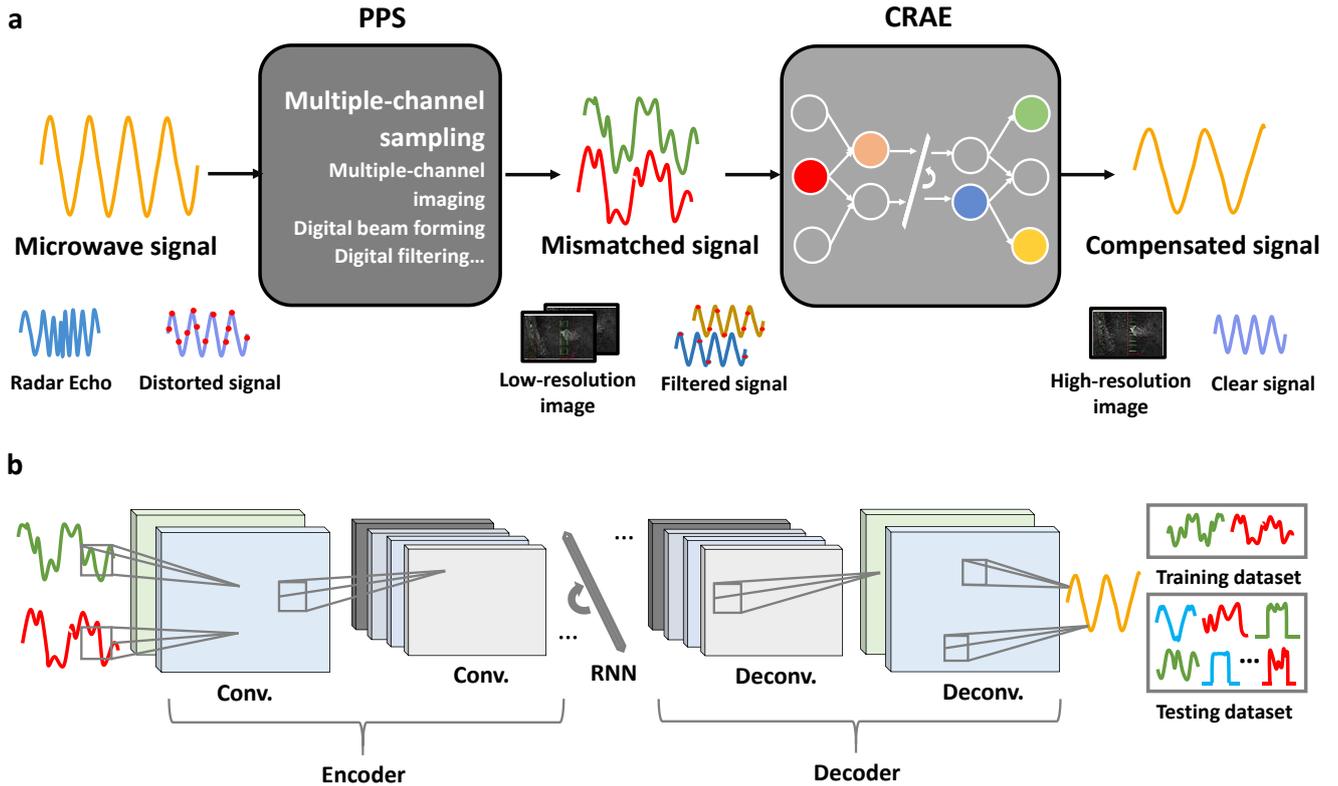

**Fig. 1| Schematic representations of hardware-irrelevant PPSs. a,** A hardware-irrelevant PPS is composed of a PPS and a CRAE. Various signals input different PPSs to obtain outputs in different domain such as one-dimension waveforms and two-dimension images. The outputs of PPSs are distorted or worse than nominal ones due to the bad influence brought by hardware deviations. The CRAE is adopted to recover the distorted signals into nominal ones. **b,** Conv. and Deconv.: the convolution and deconvolution layer, respectively. The CRAE is composed of CNN, RNN, and an AE structure. The CRAE is trained with a category of signals collected when the PPS is in two hardware deviations which is represented by different colors. The trained CRAE can be tested on other categories of signals represented by different shapes which are collected when the PPS is in multiple fixed hardware deviations.

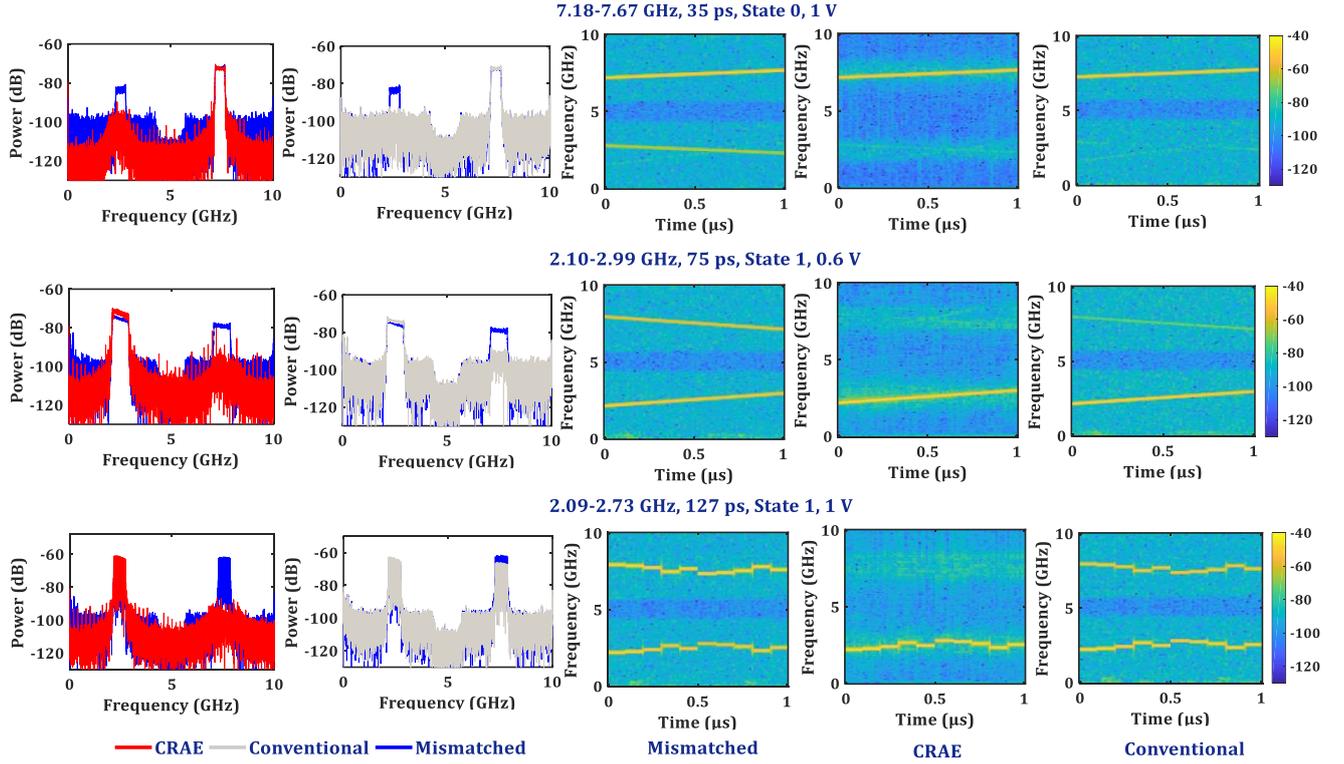

**Fig. 2 | Mismatch compensation by the CRAE compared with conventional techniques.** A mismatched LFM signal of 1 V with 35 ps under State 0 is compensated with CRAE and traditional approaches, displayed in the first row. Another two examples, a mismatched LFM signal of 0.6 V with 75 ps and a mismatched Costas frequency modulated signals of 1 V with 127 ps under State 1, are displayed in the second and third rows, respectively. In the first and second columns: the red and gray curves represent the outputs of the CRAE and conventional approaches in the FFT spectra domain, respectively; the blue curves represent the mismatched signals. The third to fifth columns: the mismatched signals, outputs of the CRAE, and outputs of the conventional method in the STFT spectra domain, respectively.

**Capability verification of the trained CRAE.** Figure 2 displays mismatched signals before and after mismatch compensation in the Fast Fourier Transform (FFT) and the Short Time Fourier transform (STFT) spectra domain. For an LFM signals of 1V with trained mismatches of 35 ps under State 0, an LFM signals of 0.6 V with untrained mismatches of 75 ps under State1, and a Costas frequency modulated signal of 1 V with untrained mismatches of 127 ps under State 1, there are mismatch components of which the frequency equals fs/2-f where fs and f are the sampling rate and frequencies of signal components, respectively. We use the trained CRAE compared with a conventional approach[15] to compensate for mismatches. As shown in the second column of Fig. 2, for the conventional method: when the volume of mismatches is small such as 35 ps, it compensates for mismatches effectively; its effects on mismatch compensation are in negative relation to volume of mismatches; it deteriorates noise floor. However, mismatch components in the outputs of the trained CRAE are negligible. Moreover, the powers of signal components are larger after mismatch compensation. Additionally, the noise floors of the outputs are lower than that of mismatched signals due to AE's strong ability to denoise. Some spurious tones in the outputs of the CRAE are caused by data segment during data processing (as analyzed in Discussion). The operation of data segment is adopted for two reasons. For one thing, our personal computer has a limited performance. For another, it is essential to improve the learning efficiency as well as to reduce computational complexity of the training processing. There are severer spurious tones when the signal accessing into the parallel photonic sampling system is of a lower amplitude. The results shown in Fig. 2 verify the CRAE's excellent ability of mismatch compensation as well as its generalizability and robustness. More results are displayed in Supplementary Fig. 3 and 4.

In fact, deep neural networks can automatically extract abstract, critical, and high-dimension features from data and acquire the latent and intricate laws. Moreover, according to the universal approximation theorem[20, 21], neural networks can develop the ability to approximate arbitrary complex functions through training. However, the majority of neural networks in computer vision[22], medical diagnosis[23], speech separation[24], and other fields such as system design[25-28] and system performance improvement[19, 29-31] are ineffective on untrained features, objects, or scenes. While the minority of neural networks can be applied in different scenes when networks are retrained according to a certain scene with the same architecture and hyperparameters[32]. They still cannot work on data with untrained features and objects. This is consistent with the "No Free Lunch" theory[33] in machine learning, which states that networks can learn latent laws from data and that any network is effective in a limited input domain. Moreover, the capacity of network depends on network types, network architectures, and hyperparameters setting. For most networks, they only acquire features of data and thus only applicable to input-like data. Under the above condition, the training data collection is cumbersome and becomes a huge burden as numerous samples would be required. Nevertheless, the proposed CRAE learns system features and thus can be generalized to features such as untrained fixed hardware deviations, objects such as untrained categories of signals, and robust to random hardware deviations such as untrained system states. In terms of overcoming hardware deviations in PPSs, the proposed CRAE implementation is an initial demonstration but not the only approach to be generalized to untrained conditions and robust to random system states.

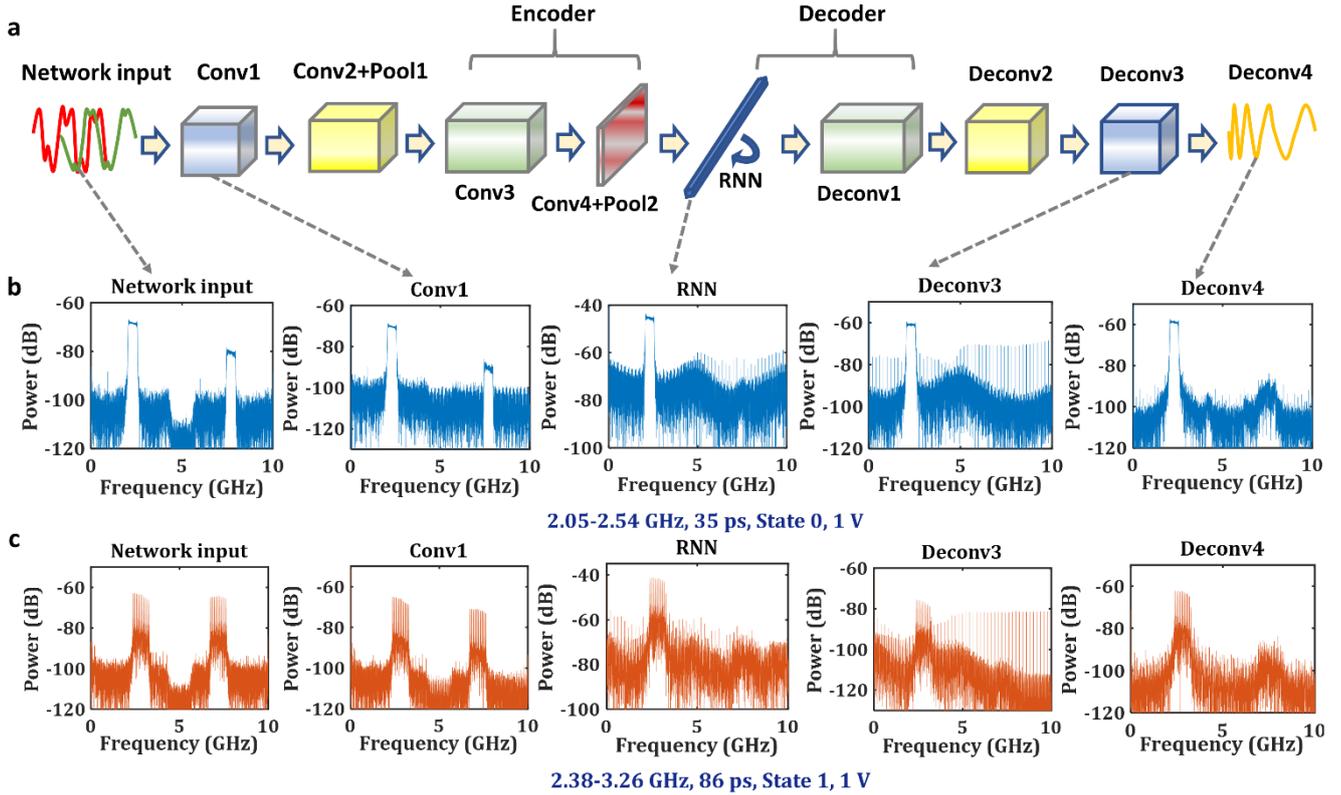

**Fig. 3 | Visualization of the mismatch compensation processing of the trained CRAE. a** The network architecture of the proposed CRAE. Conv1-Conv4: the convolutional layers; Pool1-Pool2: the max-pooling layers; RNN: the recurrent neural network layer. The signals with green and red colors in **a** represent signals with mismatches of 35 ps and 57 ps, respectively. The Deconv4 is the network output layer. **b, c** Two examples of visualization of an LFM signal of 1 V in 2.05-2.56 GHz range with mismatches of 35 ps under State 0 (as shown in blue curves) and a Costas frequency modulated signal of 1 V in 2.38-3.26 GHz range with mismatches of 86 ps under State 1 (as shown in red curves) in the FFT spectra domain.

**Interpretability of the CRAE.** To explain the working principle of the CRAE and to better understand its generalizability and robustness capability, we visualize and analyze the processing of mismatch compensation of the trained CRAE on signals under trained and untrained conditions. Two examples are demonstrated. One is an LFM signal of 1 V in 2.05-2.54 GHz range with mismatches of 35 ps under State 0 and the other is a Costas frequency modulated signal of 1 V in 2.38- 3.26 GHz range with mismatches of 86 ps under State 1. Fig. 3b and c are the visualization of the network input layer, the Deconv4 layer (i.e. network output), and the first feature map of some hidden layers in the proposed CRAE (see Fig. 3a) in the FFT spectra domain.

From the network input layer to the Conv1 layer, the power of the mismatch components in the mismatched LFM and Costas frequency modulated signals decreases. After the RNN layer, the mismatch components of both signals are almost eliminated and the signal noise in the following layers is gradually fitted. In Fig. 3b and c, during the processing of mismatch compensation, the laws of signals under untrained conditions (i.e. the above-mentioned Costas frequency modulated signal), match well with that of signals under trained conditions (i.e. the above-mentioned LFM signal). This implies that the trained CRAE has learned the inherent patterns of the parallel photonic sampling system. It also explains the reason why CRAE generalizes well to the untrained cases and remains robust to system states.

It is found that the trained CRAE acquires the inherent patterns of the parallel photonic sampling system and removes bad influence brought by fixed hardware deviations, i.e. time mismatches. Therefore, the parallel photonic sampling system integrating the trained CRAE is hardware-irrelevant whereas the traditional parallel photonic sampling system without a CRAE is hardware-relevant.

**Demonstration of the hardware-irrelevant parallel photonic sampling system.** We demonstrate the hardware-irrelevant parallel photonic sampling system. With assistance of LabVIEW, the digital output of the parallel photonic sampling system is automatically collected into a personal computer and instantly accesses into the trained CRAE after data processing. Under States 1-3, several LFM signals of various amplitudes and Costas frequency modulated signals of 1 V enter the parallel photonic sampling system in various mismatch degrees. For the hardware-relevant parallel photonic sampling system: from the time domain, the output waveforms are distorted (see the blue curves in Fig. 4a) with the distortion of waveforms being in positive relation to volume of mismatches; from the FFT and STFT spectra perspective, there are mismatch components (see Fig. 4c and blue curves in Fig. 4d). However, under whichever system states, the hardware-irrelevant parallel photonic sampling system can rapidly output smoother and more regular waveforms online. There is scarcely any mismatch component (see Fig. 4b and red curves in Fig. 4d) even when the fixed hardware deviations in parallel photonic sampling system are so large that the power of mismatch components surpasses that of the signal components (see the second and forth columns of Fig. 4c). Moreover, the power of signal components in the output of the hardware-irrelevant system under States 3 approximates that under State 1, detailed in Discussion. Note that the noise floor of the output of the hardware-irrelevant system is slightly different because of different random system states. More demonstrations are displayed in Supplementary Fig. 5-7. Hence, we can believe hardware deviations barely have any influence on the performance of the hardware-irrelevant parallel photonic sampling system.

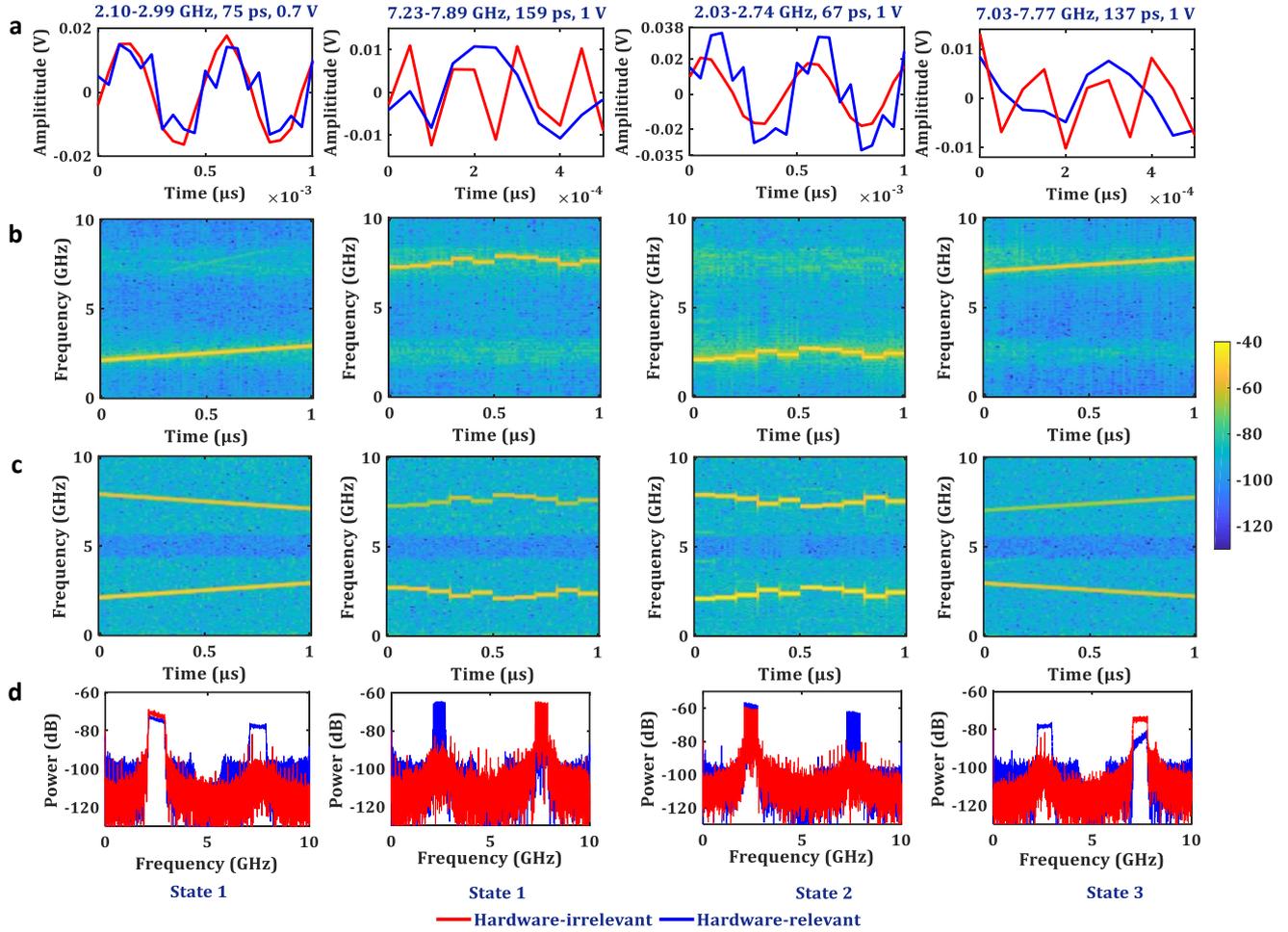

**Fig. 4 | Demonstration of the hardware-irrelevant parallel photonic sampling system. a**, In the time domain; **b, c**, In the STFT spectra domain; **d**, In the FFT spectra domain. An example of LFM signal of 0.4 V with mismatches of 75 ps under State 1 and another example of LFM signal of 1 V with mismatches of 137 ps under State 3 are displayed in the first and fourth columns, respectively. Two other examples of mismatched Costas frequency modulated signals of 1 V with mismatches of 159 ps under State 1 and with mismatches of 67 ps under State 2 are presented in the second and third columns, respectively. The red curves and blue curves are represented the outputs of hardware-irrelevant and hardware-relevant parallel photonic sampling system, respectively.

**Discussion**
By virtue of the combination of CNN, RNN, and AE structure and the proposed training approach, the CRAE demonstrates excellent generalizability and robustness. Prior to the formation of the CRAE, three other neural networks are established. The first is a convolutional autoencoder (CAE), which is the CRAE without the RNN layer. The second is an RNN, which is the RNN layer of the CRAE only. The third is a temporal convolutional network (TCN), which is the CRAE with two additional dilated convolutions behind the Pool1 and Pool2 with dilation rate of 4. We have trained these three networks using LFM signals with mismatches of 35 ps and 57 ps. The CAE and RNN are moderately effective on mismatch compensation for signals with untrained mismatch degrees, while the TCN network almost does not work (Supplementary Fig. 8). This may be because the receptive field is so large that the TCN extracts features based on data completely and has difficulty in generalizing to untrained cases. In addition, we have trained a CRAE using LFM signals with mismatches of 35 ps, and the network is effectively only on signals with mismatches of 35 ps, while leaving only a just a bit or none of effect on signals with untrained mismatches such as 92 ps (Supplementary Fig. 9). We speculate that the CRAE trained using signals in one mismatch degree tends to abstract features of data on their own. When we train a CRAE using signals in two different mismatch degrees, two signals which are different in the power of the mismatch components need to be mapped into one same signal. To achieve this goal, it is essential that the CRAE should extract the system features, i.e. time mismatches, and thus can be effective on signals with untrained features.

To analyze the reasons for spurious tones in the outputs of the CRAE, we establish a convolutional autoencoder (CAE) which is the CRAE without the RNN layer. For the training and testing datasets of the CAE, we do not operate data segment. The output of the CAE has no spurs (Supplementary Fig. 10). Hence, we can believe that the spurious tones are brought by data segment. The reason why we do not use the CAE for mismatch compensation is that the CAE cannot work well on signals with untrained mismatch degrees.

In addition, we analyze the power of signal components in the output of the hardware-irrelevant parallel photonic sampling system under States 1-3. As a result, the power of the signal components in the output of the hardware-irrelevant system approximates that under State 1. An example is demonstrated in Supplementary Fig. 11. Under State 3, an LFM signal in 2.07-2.86 GHz enters the parallel photonic sampling system with mismatch degrees of 61 ps. The power of the signal components in the output of the hardware-irrelevant system (see the red curve in Supplementary Fig. 11) is larger than its counterpart in the parallel photonic sampling system with mismatch degrees of 0 ps under State 3 (see the gray curve in Supplementary Fig. 11). Moreover, the power approximates its counterpart in the parallel photonic sampling system with mismatch degrees of 0 ps under State 1 (see the green curve in Supplementary Fig. 11). This is because that the training dataset of the CRAE is collected under State 0 and that the power of the output in the parallel photonic sampling system under State 1 is almost the same with that under State 0.

For any network is effective in a limited input domain, we explore the boundary where the trained CRAE is ineffective on mismatch compensation. For an LFM signal in 2.02-2.82 GHz range, when the volume of the mismatches is over 165 ps, some mismatch components of high frequencies cannot be compensated even with enhanced power of signal components. When the volume of mismatches is up to 220 ps, the power of signal components is extremely low and thus the CRAE is almost ineffective (Supplementary Fig. 12). Nonetheless, the trained CRAE performs better on signals with large mismatch degrees than most conventional approaches.

In summary, through integrating a CRAE, which can be generalized to features such as untrained fixed hardware-deviations and objects such as untrained categories of signals, and can robust to random hardware deviations such as untrained system states, we accomplish a hardware-irrelevant PPS (parallel photonic sampling in the experimental demonstration) where the hardware deviations are irrelevant to system performance. Our approach is extensively applicable to achieving various PPSs which are either discrete or integrated in photonic, electric, and other fields. Hence, it paves the way for performance enhancement of modern information systems and data processing such as radar, communication, and digital filtering.

## Methods

**Experimental setup of parallel photonic sampling system.** In the two-channel 20-Gsample/s photonic sampling system (Supplementary Fig. 1a), an actively mode-locked laser (AMLL, CALMAR PSL-10-TT) driven by a microwave seed (KEYSIGHT E8257D) at a frequency of 20 GHz emits optical pulses at a 20-GHz repetition rate. The optical pulse train from the AMLL is amplitude modulated by the signal to be sampled via a Mach–Zehnder modulator (MZM, PHOTLINE MXAN-LN-40) and thus the signal was sampled with a fixed interval. Before the sampled signal enters a dual-output MZM (DOMZM, PHOTLINE AX-1×2–0MsSS-20-SFU-LV), we use an analog tunable delay line (ATDL, General Photonics MDL-002) with a tuning accuracy of 1ps to allow one optical pulse of two adjacent pulses to pass through the DOMZM at its maximal transmission rate and allow the other pulse to pass through the MZM at its minimal transmission rate. The DOMZM is driven by a 10 GHz microwave signal into which a custom-built frequency divider transfers the 20-GHz signal from the microwave seed. Therefore, the optical pulse train was demultiplexed into two channels. There are two identical digital tunable delay lines (DTDL, Motorized Delay Line MDL-002) with a tuning accuracy of 1fs before two identical custom-built photodetectors (PD) of 10-GHz bandwidth. After PDs convert optical pulses into voltage, a multichannel real-time oscilloscope (OSC, KEYSIGHT DSO-S 804A) was adopted as the analog-to-digital converter (ADC), which has a 10-GS/s sampling speed and four channels. Hence, the two-channel 20-Gsample/s photonic sampling system outputs two-way digital signals. The OSC was synchronized by the microwave seed to keep the quantization clock synchronized with the AMLL. The signals to be sampled are generated via an arbitrary waveform generator (AWG, KEYSIGHT M8195A). By adjusting two DTDLs, we can change fixed hardware deviations, i.e. physical lengths, between two channels and thus obtain signals with mismatches of various values.

**CRAE architecture.** The CRAE (Supplementary Fig. 1b) includes four convolution layers which have strong ability to extract high-level features, two pooling layers which can decrease the redundant information of input data and avoid overfitting, a recurrent neural network layer which is very adaptive to deal with sequence-like objects, and four deconvolution layers for up-sampling. In addition, the CRAE embodies an encoder-decoder structure which is good at abstracting critical information from data. The first convolutional layer Conv1 convolves 32 filters of 1 × 2 with stride 1. The second convolutional layer Conv2 convolves 64 filters of 1 × 2 with stride 1 followed by a max-pooling operation Pool1 with stride 2. The third convolutional layer Conv3 convolves 128 filters of 1 × 3 with stride 1. The fourth convolutional layer Conv4 convolves 1 filter of 1 × 1, which adds all features of the Conv3 layer. Another max-pooling operation Pool2 with stride 2 is following. The Conv1, Conv3, Pool1, and Pool2 are followed by a nonlinearity activation function of Tanh. The first deconvolutional layer Deconv1 deconvolves 128 filters of 1 × 3 with stride 1. Between the Pool2 and Deconv1, we embed an RNN layer. The number of neurons in RNN layer is 128. The second to forth deconvolutional layers Deconv 2-4 deconvolve 64, 32, and 1 filters of 1 × 3 with stride 1, respectively. The Deconv4 is also the network output layer. The Conv3 to the Conv4 layer and the RNN to the Deconv1 layer can be regarded as encoder and decoder operation, respectively. Network weights are initialized with a Xavier initializer[34]. Network biases are initialized by truncated normal function with 0 mean and 0.1 stddev.

**Experimental implementation, data generation, and data acquisition.** We modulate an LFM signal to be sampled into MZM and control OSC to collect the two-way digital outputs into a personal computer via LabVIEW. We develop a LabVIEW program (Program 1) through which we modify the mismatch degrees of the parallel photonic sampling system to 0 ps or known values. In Program 1, two-channel signals collected from OSC are interleaved; the interleaved signal is transformed into frequency domain via FFT; the waveform and the FFT spectra graph is displayed. When there are mismatch components in the FFT spectra graph, we believe there are fixed hardware deviations in the two channels. We adjust the DTDLs before PDs to add the physical length of a certain channel, until there is almost no mismatch component in the FFT graph. It is found that the mismatch degrees of the parallel photonic sampling system are 0 ps. Hence, we obtain signals with mismatches of 0 ps. In addition, by adjusting the DTDLs to add various fixed delays of a certain channel, we can obtain data with mismatches of known values.

We develop another LabVIEW program (Program 2) to control AWG to emit amplitude/frequency-varying signals and to automatically collect digital data from OSC into a personal computer. In our experiment, signals to be sampled are LFM signals and Costas frequency modulated signals in 2.0-3.3 GHz and 7.0-8.3 GHz range. The sampling rate of the AWG and the OSC are 60 and 10 GSa/s, respectively.

We develop the third LabVIEW program (Program 3) to test the performance of the hardware-irrelevant parallel photonic sampling system. In Program 3, the two-way data sampled by OSC with 20,000 points each way, is instantly collected into personal computer and inaccurately cut to retain signal segment with 10,000 points each way; after interleaving, the interleaved signal with 20,000 points is separated into 200 segments and input into a trained CRAE model; then, the network output, i.e. the output of the hardware-irrelevant system, is displayed in time and FFT spectra domain to intuitively real-time observe the performance of the hardware-irrelevant parallel photonic sampling system. System output is saved.

We conduct four experiments (Experiments 0-3) in four different days with four random system states (States 0-3). For different random system states, state of each hardware element of the parallel photonic sampling system, such as microwave instruments and optic-electronic devices, is different due to variant environment, random started noise, etc.

In Experiment 0, the corresponding random system states is regarded as State 0. First, with the assistance of Program 1, we modify the mismatch degrees of the parallel photonic sampling system to 0 ps. Second, with the assistance of Program 2, we control AWG to produce 6000 1μs LFM signals which are modulated into the optical pulses via MZM and control OSC to automatically collect two-way system outputs. Those signals are of 1 V in 2.0-3.3 GHz or 7.0-8.3 GHz range, each range 3000. Note that we set a time delay of 500 ms between each signal to ensure the digital output of the former signal is collected before the latter signal starts generate. Hence, we obtain 6000 LFM signals with mismatches of 0 ps. The entire time of above data collection is about 25 minutes. Third, by adjusting the DTDL to add delays of 35 ps, 57 ps, 79 ps, and 92 ps in a certain channel, the mismatch degrees of the parallel photonic sampling system are 35 ps, 57 ps, 79 ps, and 92 ps. Via Program 2, we obtain LFM signals with mismatches of 35 ps, 57 ps, 79 ps, and 92 ps. Those signals are adopted to train and test the trained CRAE.

In Experiment 1, the corresponding random system states is regarded as State 1. Under State 1, the power of the output of the two-channel 20-Gsample/s photonic sampling system is almost the same with that under State 0. First, with the assistance of Program 1, we modify the mismatch degrees of the parallel photonic sampling system to 0 ps. Second, with the assistance of Program 2, we control AWG to produce 400 1 μs LFM and 400 1 μs Costas frequency modulated signals which are modulated into optical pulses via the MZM and control OSC to automatically collect data. Those LFM and Costas frequency modulated signals are in 2.0-3.3 GHz or 7.0-8.3 GHz range. The amplitudes of LFM and Costas frequency modulated signals are various values and 1 V, respectively. Hence, we obtain LFM and Costas frequency modulated signals with mismatches of 0 ps. Third, we adjust the DTDL to add delays from 30 ps to 160 ps in a certain channel and via Program 2, we obtain LFM and Costas frequency modulated signals with mismatches from 30 ps to 160 ps. Those mismatched signals are used to test the generalizability and robustness of the trained CRAE. In addition, with the assistance of Programs 2, we use AWG to randomly produce a 1 μs LFM or Costas frequency modulated signal of 1 V in 2.0-3.3 GHz or 7.0-8.3 GHz range. The signal to be sampled is modulated into the MZM when the mismatch degrees of the parallel photonic sampling system are 30 ps to 160 ps. Through Program 3, the digital output is interleaved, cut, and then instantly input into the trained CRAE model. We observe the network output and evaluate the performance of the hardware-irrelevant parallel photonic sampling system.

Experiments 2 and 3 aim to test the hard-irrelevant system performance. In Experiment 2, the corresponding random system state is regarded as State 2. Under State 2, the power of the output of the two-channel 20-Gsample/s photonic sampling system is higher than that under State 0. With the assistance of Program 2, we control AWG to randomly produce a 1 μs Costas frequency modulated signal of 1 V in 2.0-3.3 GHz or 7.0-8.3 GHz range. The signal to be sampled is modulated into the MZM when the mismatch degrees of the parallel photonic sampling system are 30 ps to 160 ps. The digital output is interleaved, cut, and instantly input into the trained CRAE model with the assistance of Program 3. We observe the network output and evaluate the performance of the hardware-irrelevant parallel photonic sampling system.

In Experiment 3, the corresponding random system states is regarded as State 3. Under State 3, the power of the output of the two-channel 20-Gsample/s photonic sampling system is lower than that under State 0. With the assistance of Program 2, we use AWG to randomly produce 1 μs LFM signals of 1 V in 2.0-3.3 GHz or 7.0-8.3 GHz range. The signal to be sampled is modulated into the MZM when the mismatch degrees of the parallel photonic sampling system are 30 ps to 160 ps. The digital output is interleaved, cut, and instantly input into the trained CRAE model with the assistance of Program 3. We observe the network output and evaluate the performance of the hardware-irrelevant parallel photonic sampling system.

**Data processing.** All digital signals collected in Experiments 0-3 from OSC are two-way and the signal in each way is with 20,000 points including non-signal segments. Because the sampling rate of the AWG and the OSC are 60 and 10 GSa/s, respectively, we perform down-sampling with factor 6 on the corresponding original data created by AWG to obtain down-sampling data with 10,000 points each way. For the collected data to be adopted to train the CRAE and to test the performance of the trained CRAE in Experiments 0 and 1, we perform cross-correlation between data from OSC (each way data with 20,000 points) and the down-sampling data (each way data with 10,000 points). Hence, we obtain two-way data with 10,000 points each way. We interleave the two single channels data to obtain the interleaved signal with 20,000 points. In addition, neurons of the RNN layer are fully connected. Hence, when the length of the input into the RNN layer is large, the network calculation amount is very large. Limited by performance of our computer, we separate the whole interleaved signals into 200 segments, namely each segment with 100 points. At the same time, operation of data segment also improves the learning efficiency and reduces computational complexity of the training process. For the data to be adopted to test the performance of the hardware-irrelevant parallel photonic sampling system, the two-way data collected from OSC is instantly collected and inaccurately cut to retain signal segment with 10,000 points each way. Then, the two single channels data is interleaved followed by operation of data segment. Note that we do not operate cross-correlation during the testing processing of the hardware-irrelevant system.

**Establishment of training and testing datasets.** For training and testing the CRAE, we need to build training and testing datasets. In Experiment 0, we get 6000 signals with mismatches of 35 ps and 6000 signals with mismatches of 57 ps. We randomly select 2800 LFM signals with mismatches of 35 ps and 2800 LFM with mismatches of 57 ps in both 2.0-3.3 GHz and 7.0-8.3GHz, together 11200 samples. We pair those data with corresponding signals with mismatch of 0 ps (as reference data) to form the training dataset. The left 800 samples are divided into the testing dataset of the CRAE which is named as Testing Dataset 0. In Testing Dataset 0, there are signals with mismatches of 79 ps and 92 ps in addition to 35 ps and 57 ps. In addition, data collected in Experiment 1 to be used to test the trained CRAE are regarded as the second testing dataset of the CRAE which is named as Testing Dataset 1 where includes LFM and Costas frequency modulated signals with mismatches of 30 ps to 160 ps.

**CRAE training and testing.** We use an Adam optimizer[35] with the learning rate of 0.001 as the minimization algorithm to train the CRAE network. During the training processing, the minimization algorithm aims to reduce the training loss obtained by calculating the absolute error between the network output of training dataset and reference data. Thus, network parameters are updated constantly with the training epochs to map the network input into network reference. The total training epochs is set to 20, 000 to ensure that network parameters have sufficiently converged and are not overfit. All codes use Python version 3.60. and TensorFlow framework version 1.10.0 (Google) on a server (GeForce GTX TITAN X GPU and Intel(R) Xeon(R) CPU X5570 at 2.93 GHz with 48 GB RAM about 1.5 h.

We use Testing Datasets 0 and 1 to test the performance of the trained CRAE on mismatch compensation. The Testing Dataset 0 under State 0 is to test CRAE's generalizability for untrained mismatch degrees and categories of signals. The Dataset Testing 1 under State 1 is to test CRAE's generalizability for untrained amplitudes and robustness to random system states. The data in Testing Datasets 0 and 1 input into the trained CRAE and then we analyze network outputs in the time, FFT spectra, and STFT spectra domain. In addition, we also adopt a conventional approach to compensate mismatch. The results are demonstrated in Fig. 2. More cases of mismatch compensation through the CRAE and conventional approach are shown in Supplementary Fig. 3 and 4.

**Testing of the hardware-irrelevant parallel photonic sampling system.** In Experiments 1-3, we test hardware-irrelevant parallel photonic sampling system online under different system states (States 1-3). With the assistance of Program 2, we use AWG to randomly produces a 1-V LFM signal or Costas frequency modulated signal within 2.0-3.3 GHz or 7.0-8.3 GHz range. The signal is modulated into the MZM. Through Program 3, the two-way digital signals are interleaved, cut, and instantly input into the trained CRAE model. We observe the waveform and FFT spectra of the system output in a real-time manner and roughly assess the performance of the hardware-irrelevant parallel photonic sampling system. The results of testing on hardware-irrelevant system are demonstrated in Fig. 4. More cases of system testing under States 1-3 are shown in Supplementary Fig. 5 and 6.

**Acknowledgements:** We thank Shi Hui of Intelligent Microwave Lightwave Integration Innovation Center (iMLic) for her assistance in revising English writing. Funding: supported by National Key R&D Program of China (Program No. 2019YFB2203700) and National Natural Science Foundation of China (Grant No. 61822508 and 61571292). **Author contributions:** X. Z., S. X., A. D., J. C., and R. W. contributed to the experiments; X. Z. processed the data, proposed the CRAE architecture and training method, and worked on the neural network python code; X. Z., S. X., and W. Z. prepared the manuscript; A. D. operated the conventional method on mismatch compensation. W. Z. initiated and supervised the research. **Competing interests:** The authors declare no competing interests. **Data and materials availability:** All data are available in the main text or the supplementary materials.

Supplementary Information

# Hardware-irrelevant parallel processing system

Xiuting Zou[‡], Shaofu Xu[‡], Anyi Deng, Rui Wang, Weiwen Zou[*]

State Key Laboratory of Advanced Optical Communication Systems and Networks, Intelligent Microwave Lightwave Integration Innovation Center (iMLic), Department of Electronic Engineering, Shanghai Jiao Tong University, Shanghai 200240, China. [‡]These authors contributed equally to this work. [*]Corresponding author: wzou@sjtu.edu.cn

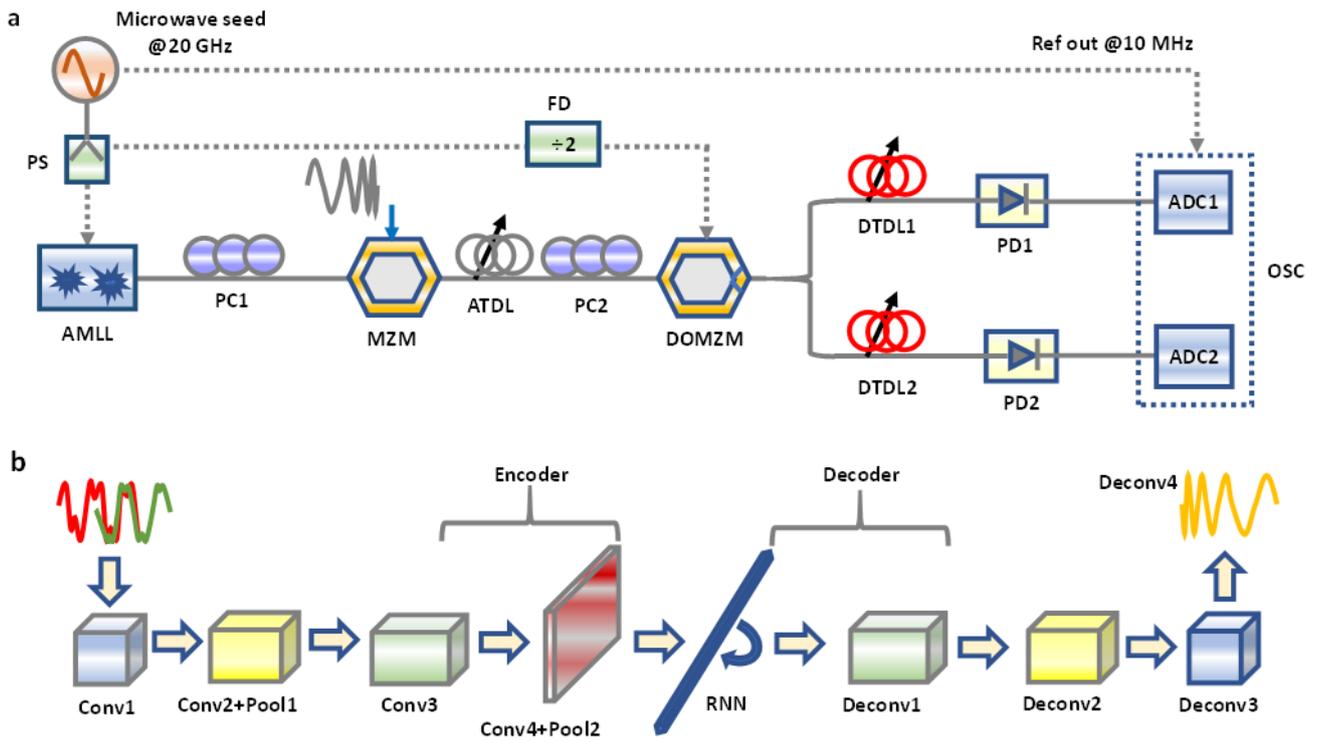

**Supplementary Figure 1 | Architecture of hardware-irrelevant parallel photonic sampling system.** This system combines a two-channel 20-Gsample/s photonic sampling system (**a**) and a CRAE (**b**). PS: power splitter; FD: frequency divider; AMLL: active mock-locked laser; PC: polarization Controller; MZM: Mach–Zehnder modulator; ATDL: analog tunable delay line; DOMZM: Dual-Output Mach–Zehnder modulator; DTDL: digital tunable delay line; PD: photodetector; ADC: analog-to-digital converter; OSC: oscilloscope. Conv1-Conv4: the convolutional layers; Pool1-Pool2: the max-pooling layers; RNN: the recurrent neural network layer. The curves with green and red colors in **b** represent signals with mismatches of 35 ps and 57 ps, respectively. The Deconv4 is the network output layer.

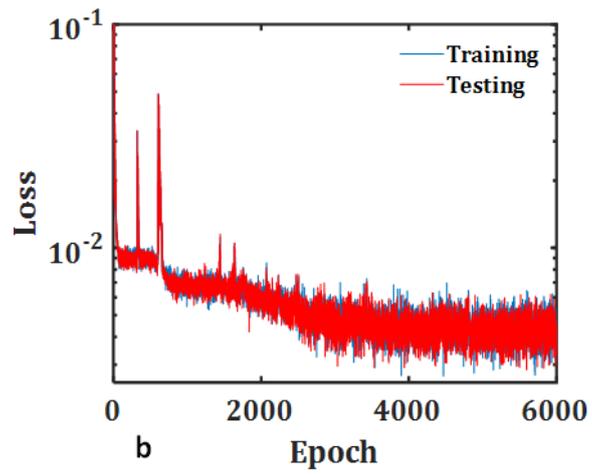

**Supplementary Figure 2 | Training and testing loss curves.** The loss is calculated by the absolute error between network output and reference data. Both training and testing loss decrease with the training epochs.

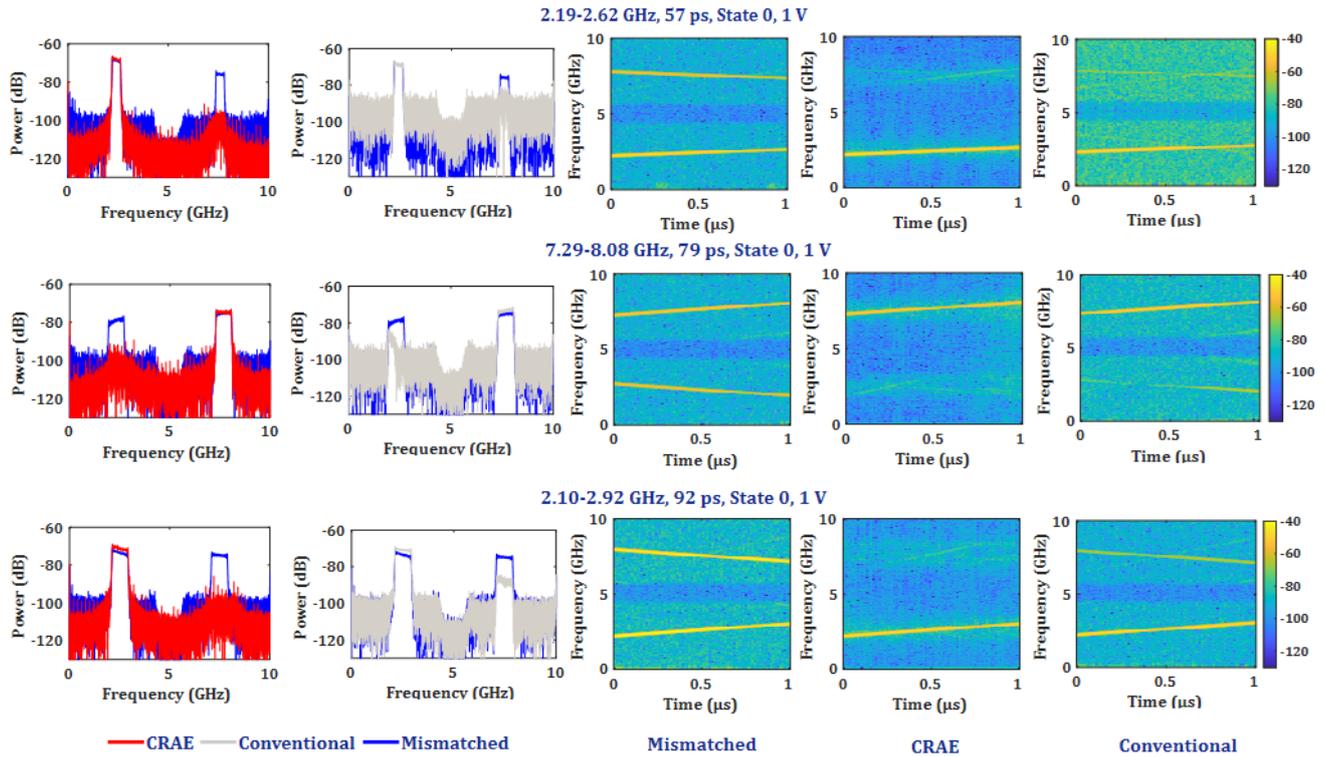

**Supplementary Figure 3 |Mismatch compensation by the CRAE compared with conventional techniques under State 0.**
Three mismatched LFM signals of 1 V with mismatches of 57 ps, 79 ps, and 92 ps under State 0, are compensated for with the CRAE and conventional approaches. In the first and second columns: the red and gray curves represent the results of CRAE and conventional approaches, respectively; the blue curves represent mismatched signals. The third to fifth columns: the mismatched signals, outputs of the CRAE, and outputs of the conventional method in the STFT spectra domain, respectively.

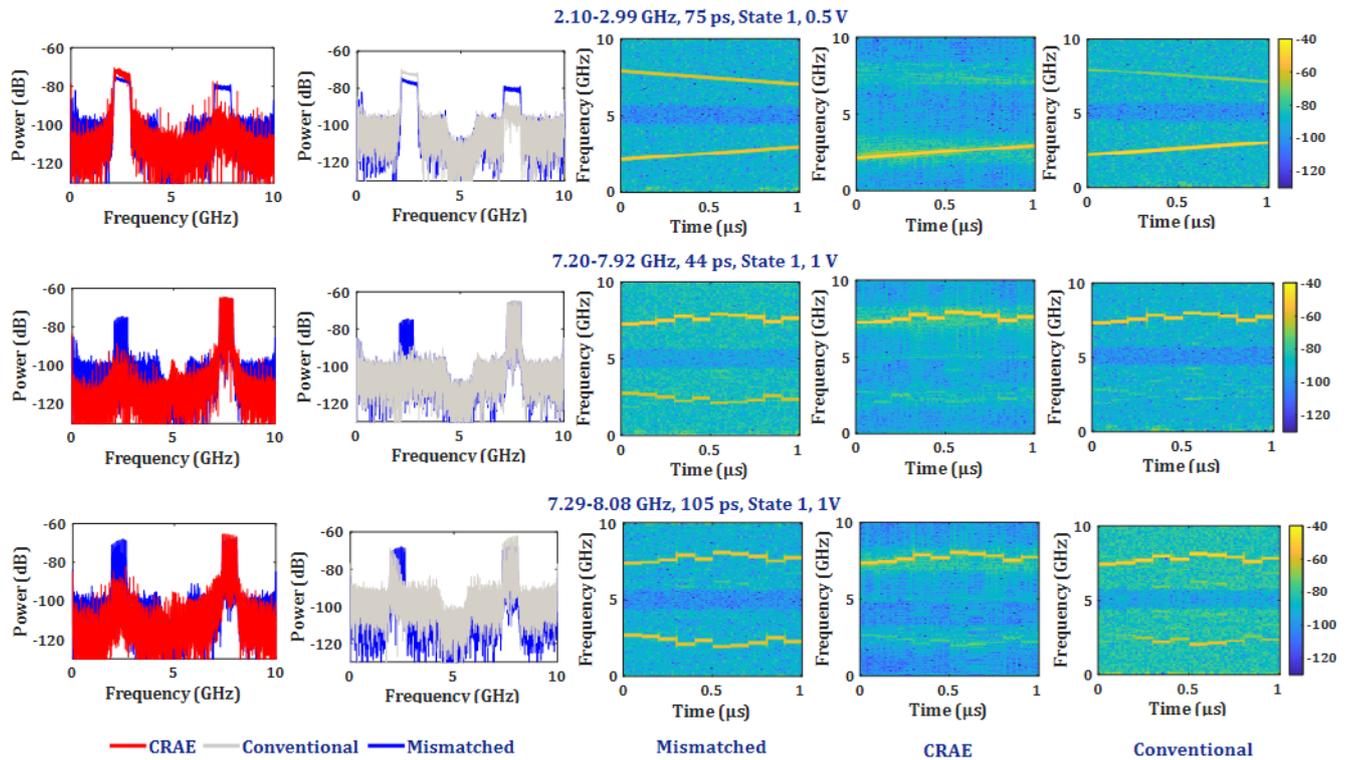

**Supplementary Figure 4 | Mismatch compensation by the CRAE compared with conventional techniques under State 1.** An LFM signal of 0.5 V with mismatches of 75 ps under State 1 is compensated for by CRAE and conventional approaches, as displayed in the first row. Two Costas frequency modulated signals of 1 V with 44 ps and 105 ps under State 1 are displayed in the second and third rows, respectively. In the first and second columns: the red and gray curves represent the results of the CRAE and traditional approaches respectively; the blue curves represent mismatched signals. The third to fifth columns: the mismatched signals, outputs of the CRAE, and outputs of the conventional method in the STFT spectra domain, respectively.

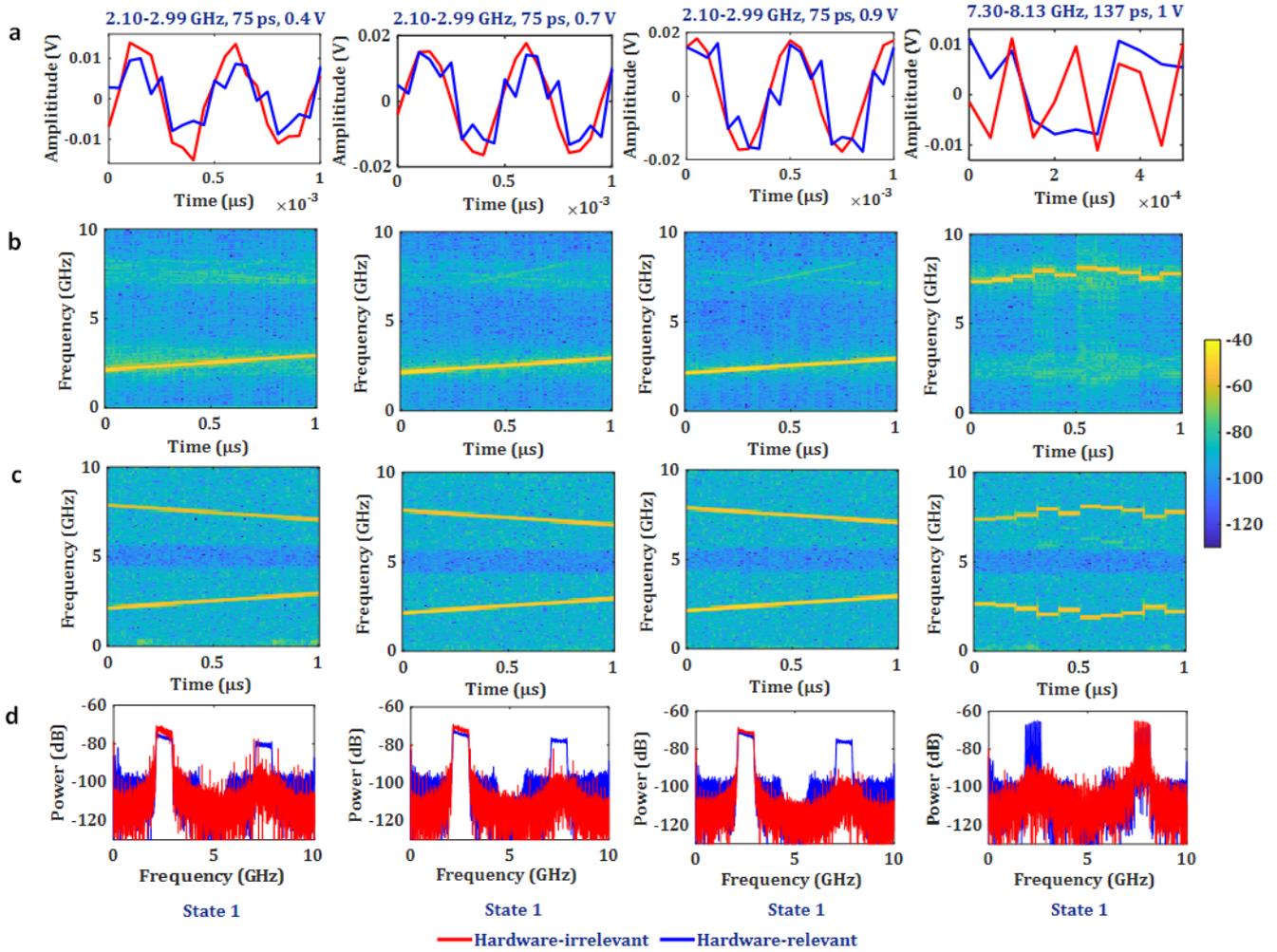

**Supplementary Figure 5 | Demonstration of the hardware-irrelevant parallel sampling system under State 1. a** In the time domain; **b**, **c** In the STFT spectra domain; **d** In the FFT spectra domain. Three examples of mismatched LFM signals of 0.4 V, 0.7 V and 0.9 V with mismatches of 75 ps under State 1 are displayed in the first to third columns respectively. Another mismatched Costas frequency modulated signals of 1 V with mismatches of 137 ps under State 1 is presented in the fourth columns. The red curves and blue curves represent the outputs of the hardware-irrelevant and the hardware-relevant parallel photonic sampling system, respectively.

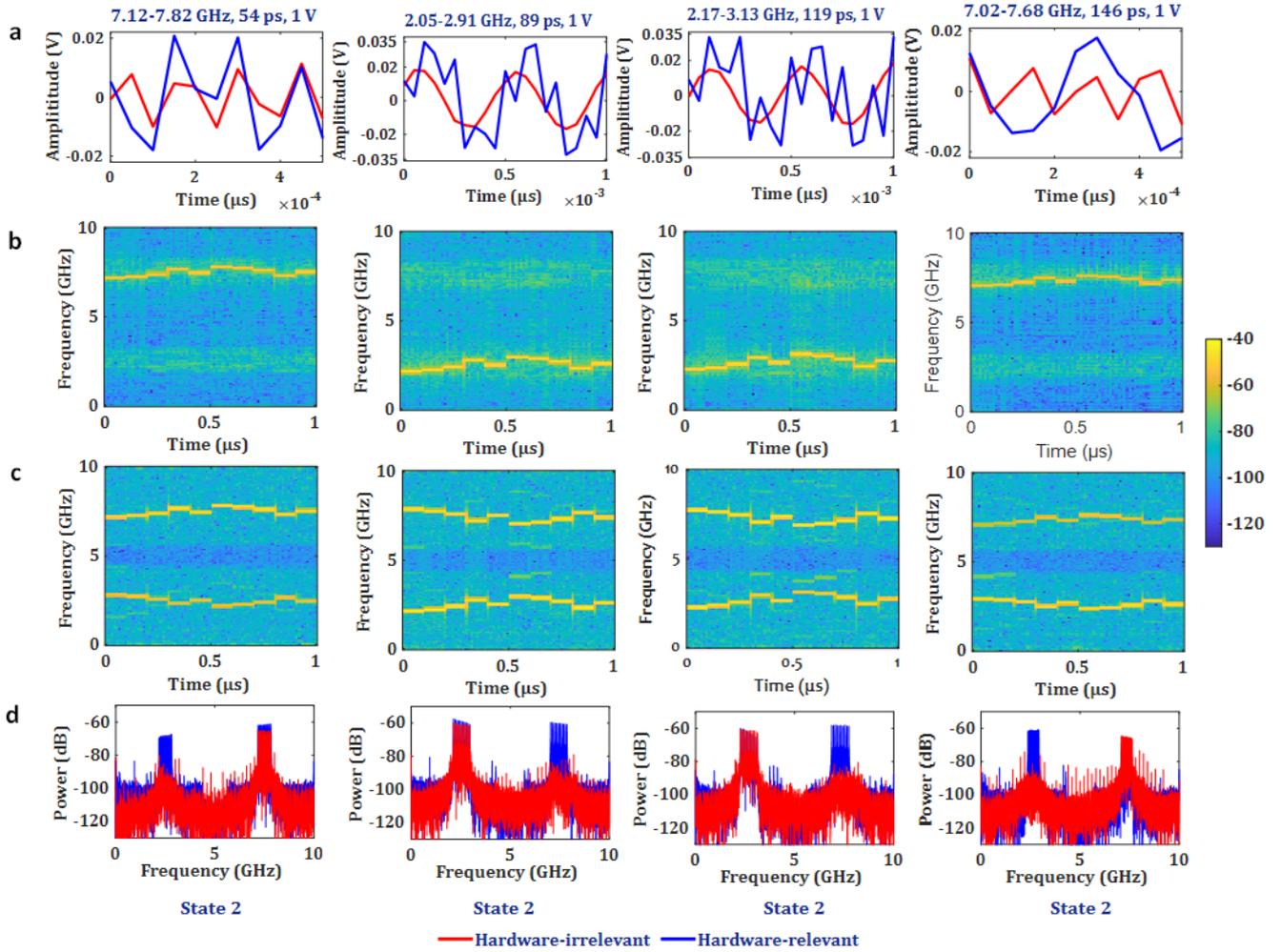

**Supplementary Figure 6 | Demonstration of the hardware-irrelevant parallel sampling system under State 2. a** In the time domain; **b**, **c** In the STFT spectra domain; **d** In the FFT domain. Four examples of mismatched Costas frequency modulated signals of 1 V with mismatches of 54 ps, 89 ps, 119 ps, and 146 ps under State 2 are displayed in the first to fourth columns, respectively. The red curves and blue curves represent the outputs of the hardware-irrelevant and the hardware-relevant parallel photonic sampling system, respectively.

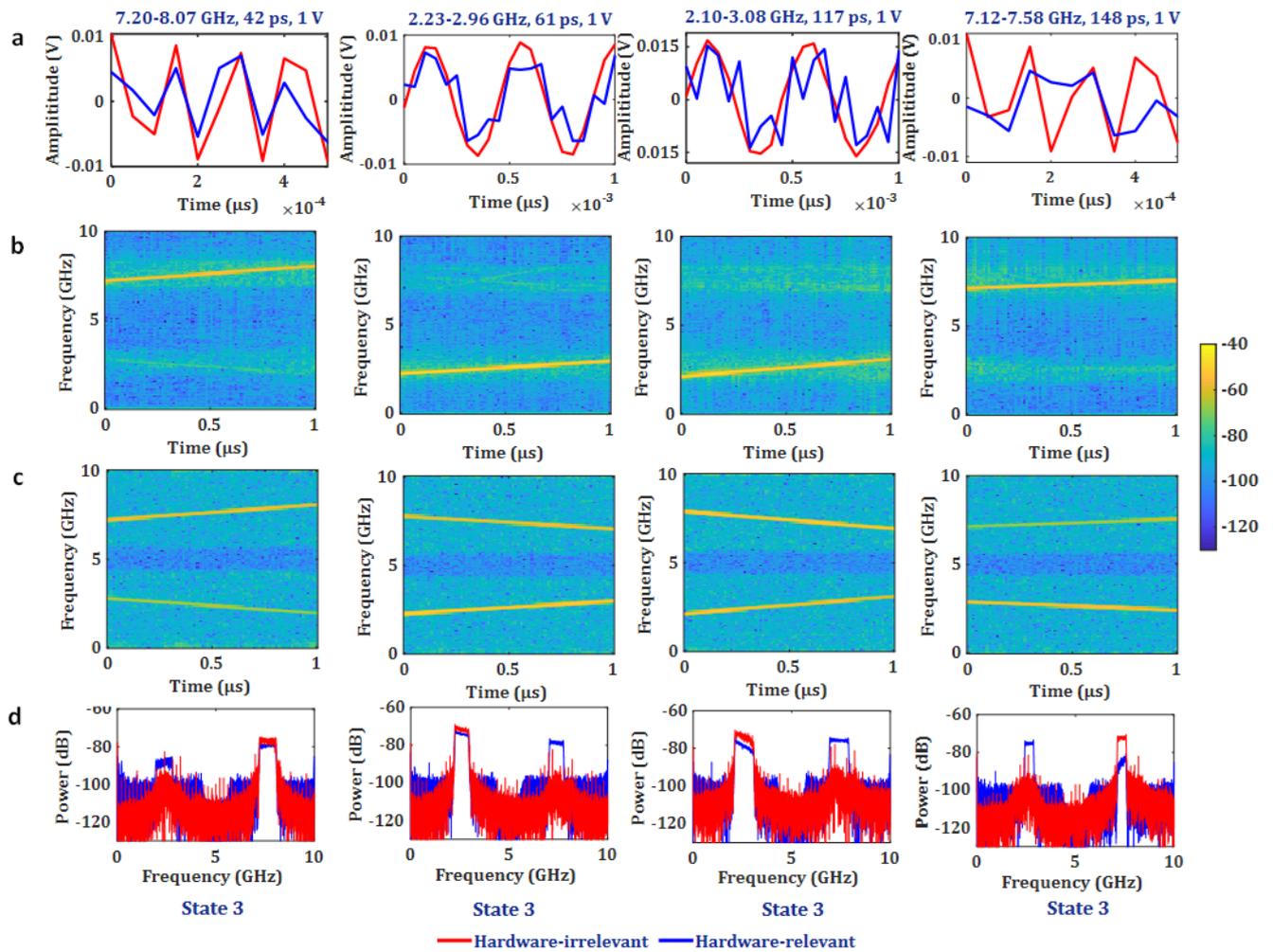

**Supplementary Figure 7 | Demonstration of the hardware-irrelevant parallel sampling system under State 3. a** In the time domain; **b**, **c** In the STFT spectra domain; **d** In the FFT spectra domain. Four examples of mismatched LFM signals of 1 V with mismatches of 42 ps, 61 ps, 117 ps and 148 ps under State 3 are displayed in the first to fourth columns, respectively. The red curves and blue curves represent the outputs of the hardware-irrelevant and the hardware-relevant parallel photonic sampling system, respectively.

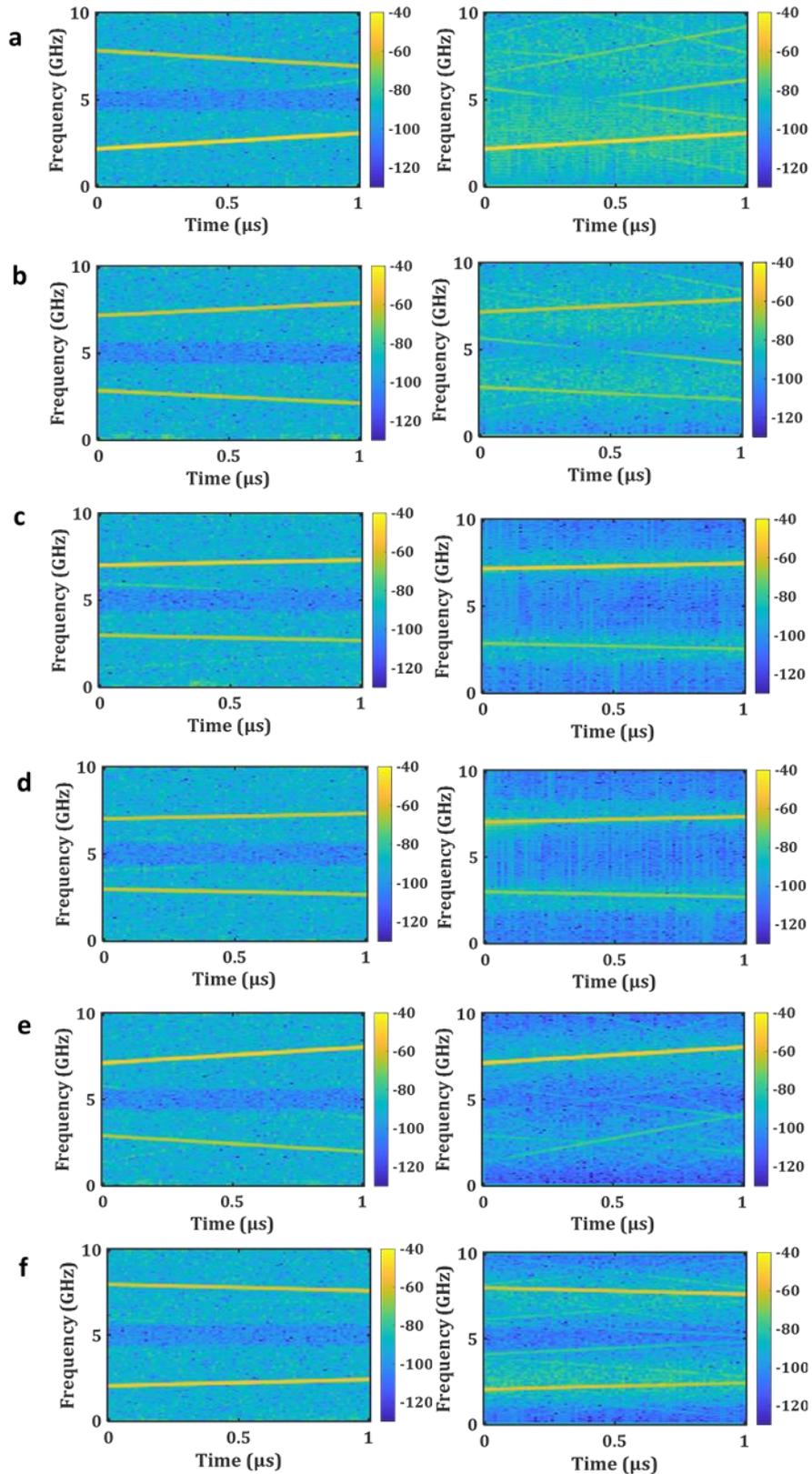

**Supplementary Figure 8 | The effects of the three networks trained using signals with mismatches of 35 ps and 57 ps.** The first and second columns are network input and output, respectively. **a**, **b** Effects of the CAE on signals with mismatches of 35 ps and 76 ps respectively; **c**, **d** Effects of the RNN on signals with mismatches of 35 ps and 76 ps respectively; **e**, **f** Effects of the TCN on signals with mismatches of 35 ps and 76 ps.

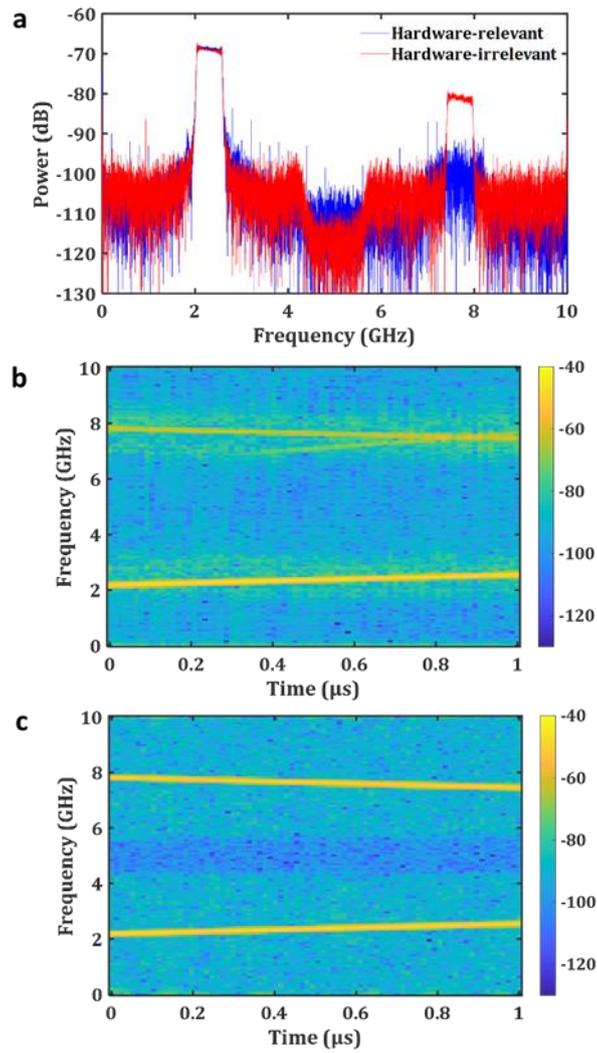

**Supplementary Figure 9 | The effect of the CRAE trained using LFM signals with mismatches of 35 ps only. a** The outcome of an LFM signal with mismatches 35 ps in the FFT spectra domain. **b**, **c** The outcome of an LFM signal with mismatches of 92 ps in the STFT spectra domain.

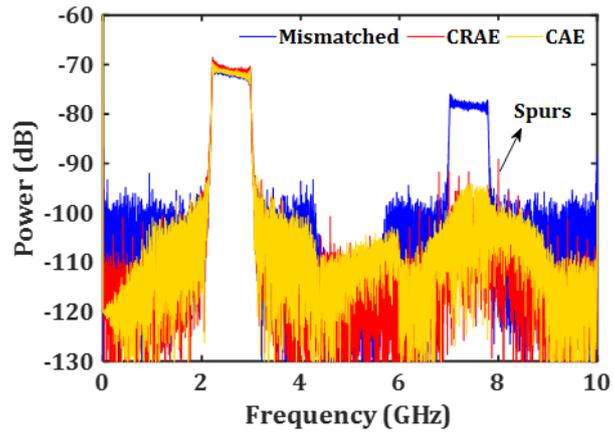

**Supplementary Figure 10 | The effect of the CRAE and CAE.** The red curve with spurious tones represents the output of the CRAE with dimension of 20,000*1. The yellow curve without spurs represents the output of the CAE with dimension of 100*200. The blue curve is the input of both the CAE and CRAE. For the CAE, we do not operate data segment during data processing.

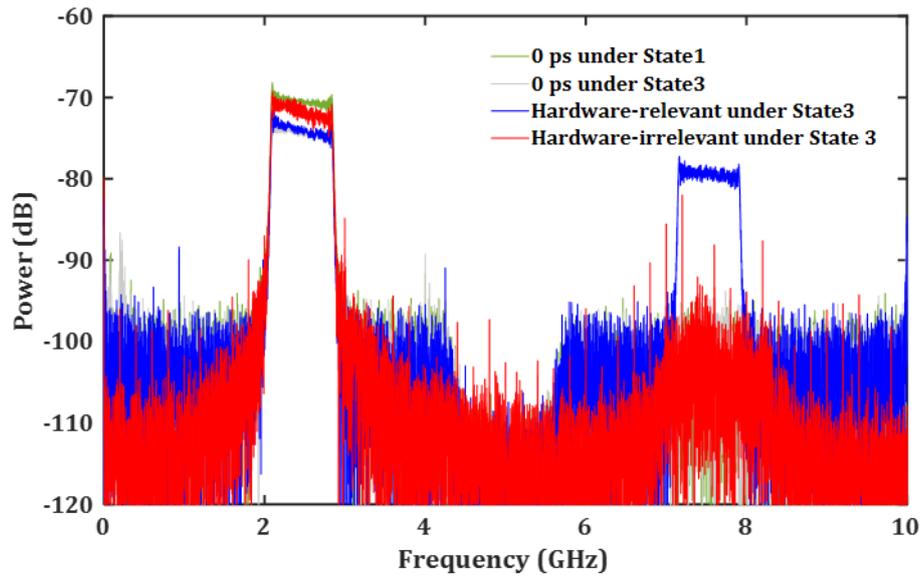

**Supplementary Figure 11 | Analysis of the power of signal components in the output of hardware-irrelevant parallel photonic sampling system.** The red and blue curves represent the output of the hardware-irrelevant system under State 3. The blue curve represents the output of the parallel photonic sampling system with mismatch degrees of 0 ps and 61 ps under State 3, respectively. The green curve represents the outputs of the parallel photonic sampling system with mismatch degrees of 0 ps under State 1. The power of the signal components in the output of the hardware-irrelevant system is larger than its counterpart in the parallel photonic sampling system with mismatch degrees of 0 ps under State 3 and approximates its counterpart in the parallel photonic sampling system with mismatch degrees of 0 ps under State 1.

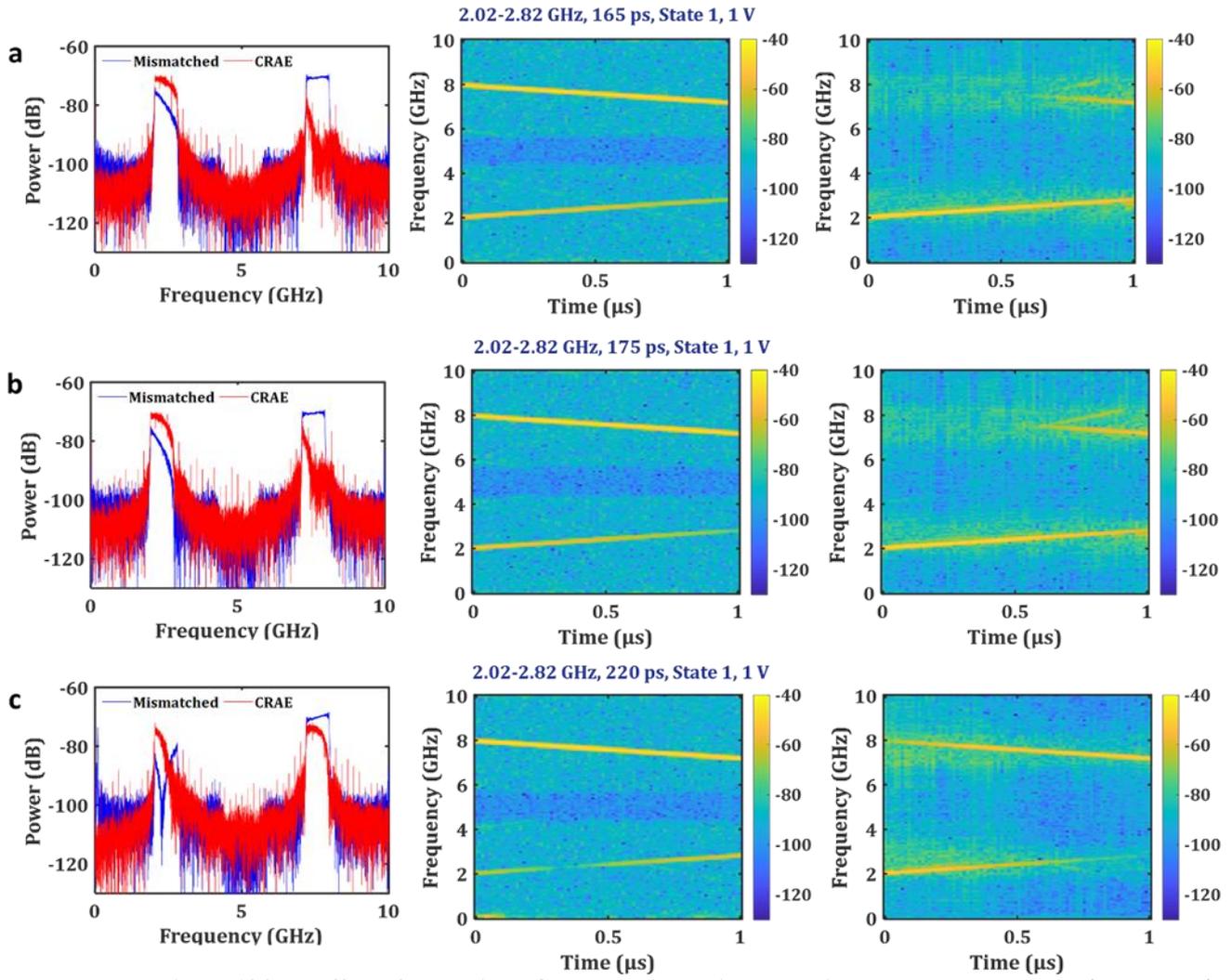

**Supplementary Figure 12 |The effect of the trained CRAE on signals in large mismatch degrees**. **a, b, c** Three examples of 1-V LFM signals in 2.02-2.82 GHz range with mismatches of 165 ps, 175 ps, and 220 ps under State 1, respectively.